# Microwave noise downconversion in interband cascade laser frequency combs

GRZEGORZ GOMÓŁKA,[1,*] FLORIAN PILAT,[2] BENEDIKT SCHWARZ,[2] CHUL SOO KIM,[3] MIJIN KIM,[4] CHADWICK L. CANEDY,[3] IGOR VURGAFTMAN,[3] JERRY R. MEYER,[3] AND ŁUKASZ A. STERCZEWSKI[1]

[1]*Laser & Fiber Electronics Group, Faculty of Electronics, Photonics and Microsystems, Wroclaw University of Science and Technology, Wybrzeze Wyspianskiego 27, 50-370 Wroclaw, Poland*
[2]*Institute of Solid State Electronics E362, Technische Universität Wien, 1040 Vienna, Austria*
[3]*Optical Sciences Division, Naval Research Laboratory, Washington, DC 20375, United States*
[4]*Amentum, Hanover, MD 21076, United States*
*\*grzegorz.gomolka@pwr.edu.pl*

**Abstract:** Chip-scale semiconductor laser frequency combs offer remarkable prospects for compact and power-efficient optical sensors. For the laser to be suitable for typical comb applications, its degree of coherence must first be assessed from a microwave self-mixing signal. Unfortunately, such measurements require scarcely available high-speed photodetectors with multi-GHz bandwidths and radio-frequency electronics. However, in this work, we demonstrate a simplified approach to comb coherence assessment for interband cascade lasers based on a relationship between easily-accessible MHz-frequency (baseband) noise and the multi-GHz-frequency intermode beat note. The downconversion of microwave noise to near-DC frequencies is found to originate intrinsically from the laser, which simultaneously acts as a frequency mixer due to electrical nonlinearities and a phase-to-amplitude noise converter due to the linewidth enhancement factor. Correlation between the electrical signals is explored in both frequency and time domains. Since this phenomenon is potentially universal in semiconductor lasers, it creates a new opportunity for frequency comb characterization, which may be particularly valuable in wavelength regions where fast photodetectors have limited availability.

## 1. Introduction

Optical frequency combs (OFCs) have greatly advanced the field of metrology by serving as direct links between optical and radio frequencies [1]. The numerous applications of OFCs include precise timing [2], telecommunications [3], ranging [4], and high-resolution spectroscopy [5–7]. Arguably, the most popular OFC generation scheme employs a mode-locked laser that emits a train of pulses in time, which corresponds to an array of equidistant lines in the frequency domain [8]. Modal coherence typically relies on active or passive mode locking techniques to ensure that the intracavity modes oscillate in phase. Pulsed operation is not, however, a prerequisite for meeting the frequency comb criterion, as the only requirement is waveform periodicity. A prime example of such a source is an electrooptically-modulated continuous wave (CW) laser that produces a frequency-modulated (FM) comb. In that case, all lines are spaced by the frequency of the external generator, even though the temporal profile lacks pulsation.

Over the past decade, a different class of compact and power-efficient OFC sources has drawn considerable attention, namely, chip-scale semiconductor lasers [9–11]. These devices with self-starting comb operation and all-electrical pumping are of particular interest due to their native emission in challenging spectral regions, including the mid-wave infrared (MWIR) [12], long-wave infrared (LWIR), and even THz. While such properties are not easily obtained in microresonator combs [13,14], they have been observed in several semiconductor laser platforms, including quantum cascade lasers (QCLs) [15–17] and interband cascade lasers (ICLs) [18,19], as well as quantum well [20–22], quantum dash [23], and quantum dot laser diodes [24,25]. Unlike mode-locked lasers, these OFCs typically have a CW FM output [26], although the generation of pulses is also possible via active [27] or passive mode-locking [28] or external compression [21,29]. The mechanism of self-starting FM comb generation in a Fabry-Pérot cavity starts with spatial hole burning (SHB), which initially promotes multi-mode lasing. Next, four-wave mixing (FWM), originating from the $\chi^{(3)}$ nonlinearity, equalizes the spacing between the otherwise-dispersed longitudinal modes via injection locking [15,26,30]. The abundance of spectral regions for which this class of devices is achievable, especially in the short and mid-wave infrared ranges, provides outstanding prospects for integrated photonic systems with reduced size and complexity. In particular, dual-comb spectroscopy (DCS) using chip-sized combs will benefit from ongoing advances in this field, since more chemical compounds can be targeted simultaneously while the acquisition speed and resolution are greatly enhanced compared to other established techniques [21,31].

A key practical parameter of OFCs is coherence, *i.e.*, the degree of phase correlation between the comb lines. A key indicator (but not proof) of stable, and hence coherent, comb operation is the appearance of a strong, narrow intermode beat note (IMB) at the repetition frequency ($f_{rep}$), which arises from the collective heterodyne beating of neighboring laser modes. In particular, the magnitude, linewidth, and noise pedestal level of the IMB are measures of the comb stability. In practice, however, the observation of a narrow IMB merely indicates that *at least two modes* are locked, while information about the locking of the other modes remains unclear. The full determination of phase relationships between the modes and a broader assessment of coherence require more advanced frequency-resolved techniques such as shifted waveform interference Fourier transform spectroscopy (SWIFTS) [32]. On the other hand, the observation of a broad IMB almost certainly guarantees either noisy comb operation or multimode-like laser emission.

Ideally, the IMB is measured optically, since even when it is possible, electrical extraction of the signal is usually more susceptible to technical noise. Because $f_{rep}$ is typically of GHz order, the demands on detector response time are challenging. While fast InGaAs and GaAs detectors with multi-GHz bandwidths (BW) are commercially available in the short-wave infrared, the selection of fast detectors operating in the MWIR and LWIR is still very scarce, and their availability is generally limited to leading research facilities [33]. While quantum cascade and quantum well detectors offer remarkable speeds of tens of GHz in the LWIR [34,35], they struggle to operate at shorter wavelengths. MWIR interband cascade [36,37] and uni-travelling carrier type-II superlattice detectors [38] have typically displayed bandwidths of only a few GHz, although resonant cavity infrared detectors (RCIDs) have operated with 3 dB frequency response up to 6.7 GHz [39]. Therefore, to advance chip-scale frequency comb research at MWIR wavelengths, it is crucial, on the one hand, to search for novel optical detection solutions, and on the other to explore alternative laser characterization methods.

In this work, we explore the relationship between near-DC low RF noise (the typical frequency range for relative intensity noise (RIN) [40,41]) and the IMB of an ICL comb emitting near 3.25 μm. We describe the intrinsic frequency conversion of noise signals in the vicinity of the IMB (at ~19.28 GHz) to near-DC radio frequencies – a phenomenon discussed previously in the context of microresonator frequency combs [42]. The semiconductor laser's electrical nonlinearity is shown to induce the device to work as a frequency self-mixer with the local oscillator at $f_{rep}$, which is analogous to a microwave diode acting as an envelope detector for AM radio. Fig. 1 schematically illustrates effects of the ICL's frequency self-mixing in four comb regimes on the laser's RIN. To emphasize that frequency downconversion from the IMB to RIN is not the fundamental origin of the latter, the mechanism of frequency-to-intensity noise conversion via the linewidth enhancement factor (LEF) is also considered.

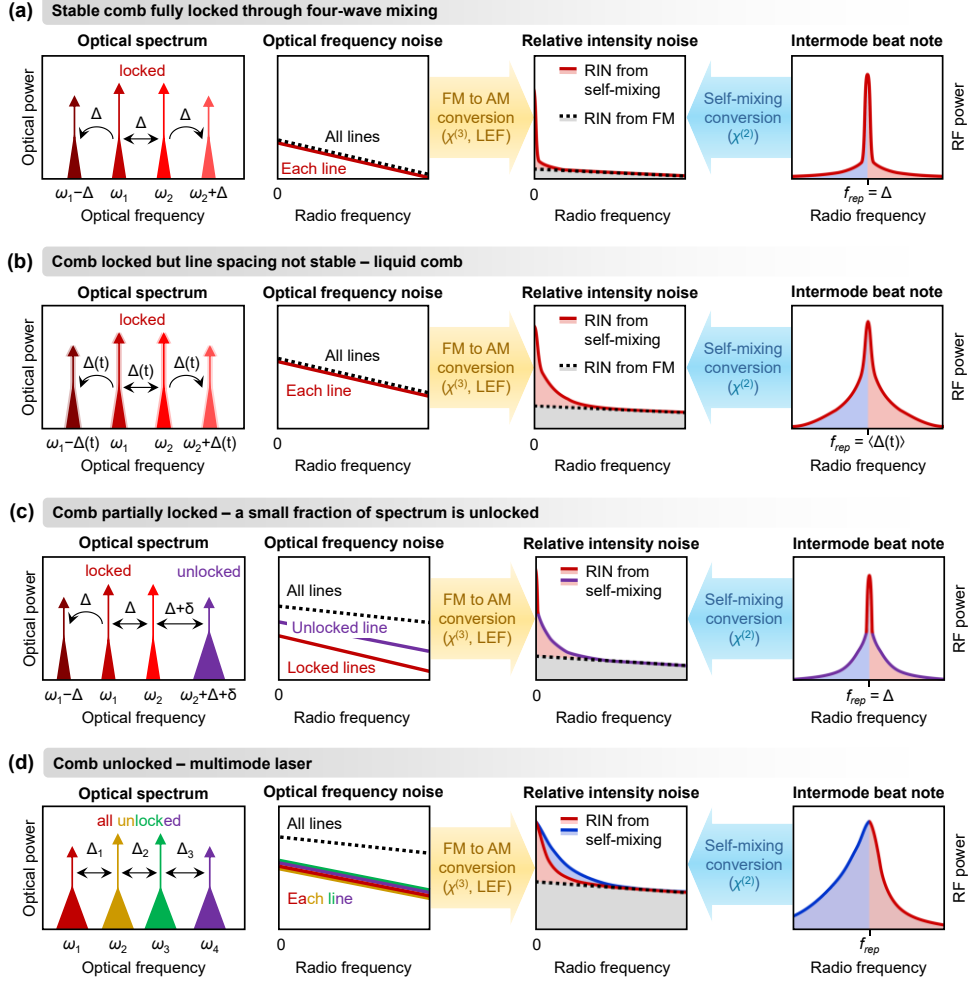

Fig. 1 Illustration of the relative intensity noise formation in ICL combs. The contributions of optical frequency noise converted to intensity noise by the linewidth enhancement factor, as well as the impact of radio frequency noise conversion from the IMB to RIN due to self-mixing, are explained. Note: frequencies higher than $f_{rep}$ (upper sidebands of IMB) downconvert in phase, while lower sidebands downconvert in counterphase due to aliasing, with the two spectra overlapping in the baseband region. This is shown schematically only in panel (d), but occurs for all the presented scenarios.

First, we lay out the theory of the noise conversion mechanism. When all the lines are fully locked and stable (Fig. 1(a)), the optical frequency noise is low and nearly equivalent to that of a single line. Due to the LEF, this frequency noise converts to RIN, albeit at a low value for the comb state. Although the strong IMB additionally converts to baseband via electrical nonlinearities (mostly $\chi^{(2)}$ resembling rectification), fortunately, there is almost no high-frequency content due to its narrow linewidth. Both contributions overlap spectrally. Next, we consider a recently-proposed liquid comb state [43] – a laser source with equidistant lines whose spacing $f_{rep}$ rapidly oscillates in time (Fig. 1(b)). In such a case, the IMB and RIN are broad and noisy due to high optical frequency noise. While the frequency noise of all lines is correlated (as expected for a comb), it is much worse than for a stable comb. Hence, it yields a noisy RIN spectrum in addition to a broad IMB. In the third regime (Fig. 1(c)) of a partially-locked state, a fraction of the optical spectrum is not locked to the remaining part of the comb. This manifests as a noise pedestal of IMB (different spacing), which then converts to pronounced RIN. The LEF has a moderate impact, since not all modes are noisy – some are noise-correlated. Finally, in an unlocked state (Fig. 1(d)), the FM-to-AM noise conversion becomes dominant as all optical modes are now uncorrelated and their noise adds cumulatively. In this case the IMB is broad (sometimes unobservable), and if strong enough, can fold to baseband. Notably, the downconversion of frequencies higher than $f_{rep}$ will occur in phase, while those lower than $f_{rep}$ occur out of phase due to aliasing. As a result, in baseband, one observes a superposition of noise for the upper and lower $f_{rep}$ sidebands.

We show below that thanks to these correlations, it becomes possible to assess the comb stability using low-frequency noise only. This enables a much simpler examination of multi-mode semiconductor lasers as potential frequency combs, especially in spectral regions where multi-GHz detectors are largely unavailable.

## 2. Experimental results

### 2.1. ICL comb device

The laser under test was an interband cascade laser grown on a GaSb substrate at the Naval Research Laboratory, which was identical to that reported in Ref. [44]. A laser bar consisting of seven Fabry-Pérot ICLs with 5 μm ridge widths, 2 mm cavity lengths, and no saturable absorbers was soldered to a copper submount, from which one was selected for the experiment. Fig. 2(a) shows a photograph of the device, along with a schematic of the measurement setup. The ICL injection current and temperature were regulated by an ultra-low-noise laser controller (Vescent Photonics, mod. D2-105-200). The light from the laser was collimated with a molded aspheric black diamond lens with anti-reflection coating and an effective focal length of 1.873 mm (Thorlabs, mod. C037TME-E). After passing through the optical isolator (Thorlabs, mod. I3400W4), the light was split into three paths. In the first, it was coupled into a 200-μm-core $InF_3$ multi-mode fiber and guided to the optical spectrum analyzer (YOKOGAWA, mod. AQ6376). Before the focusing lens for the fiber, a fraction of the light was reflected by the ZnSe window towards a highly-sensitive infrared detector (VIGO Photonics, mod. LabM-I-4) with a bandwidth of 7 MHz to create the second path. This detector, alongside the Rohde&Schwarz FPL1007 radio frequency spectrum analyzer, measured the ICL's low-frequency noise. In the third path, the laser beam was focused on an interband cascade infrared photodetector (ICIP) with a nominal bandwidth of 1.5 GHz manufactured at TU Wien [37]. The detector was reverse-biased at 4.9 V with a Sigatek SB55D4 bias tee, and its output was guided by the 1-m-long 2.92 mm microwave cable to the Rohde&Schwarz FSW RF spectrum analyzer. In this path, the IMB of the ICL comb at ~19.288 GHz was measured optically. Since all the optical and radio-frequency spectra presented throughout this paper are normalized to dark spectra, the term 'relative' optical/RF power is used.

Fig. 2(b)–(e) show the optical spectrum, CW *P-I-V* characteristics, RF IMB spectrum, and far-field beam profile for operation of the device at 20°C. The injection current was limited to 130 mA, which constrained the maximum optical output power to ~20 mW with a wall-plug efficiency of 3.9%. The non-uniform envelope of the ICL's optical spectrum and the emergence of isolated dominant modes (Fig. 2(b)) indicate potential waveguiding issues and modal leakage to the high-index GaSb substrate. This is additionally confirmed by the far-field beam profile (Fig. 2(e)), which indicates a non-negligible emission tail in the direction perpendicular to the heterostructure. Despite this, at some injection currents the laser exhibited a distinct narrow IMB as presented in Fig. 2(d).

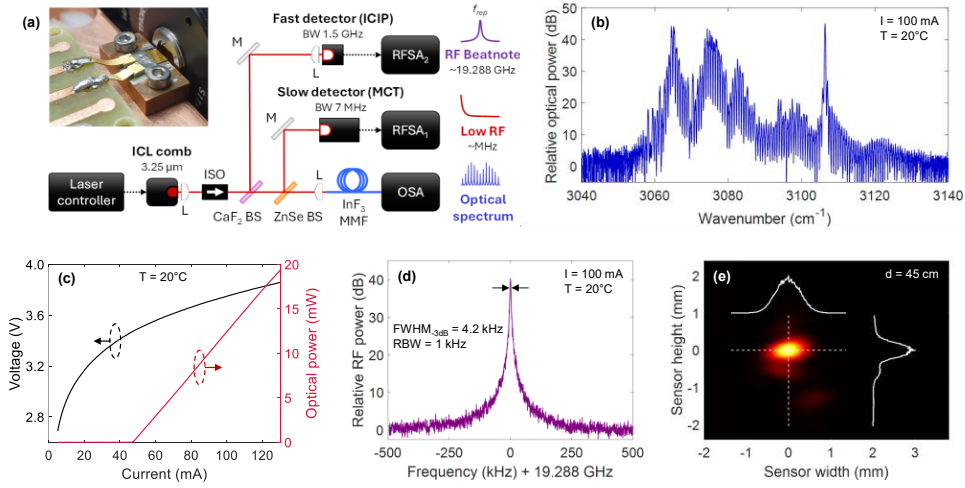

Fig. 2. (a) Photograph of the mounted ICL device and schematic of the experimental setup used for its characterization. The components are L – lens, ISO – optical isolator, BS – beam splitter, M – mirror, MMF – multimode fiber, ICIP – interband cascade infrared photodetector, MCT – mercury cadmium telluride photodetector, RFSA – radio frequency spectrum analyzer, OSA – optical spectrum analyzer. (b) Optical spectrum of the ICL comb when injected with 100 mA at $T$ = 20°C. (c) P-I-V characteristics of the laser for CW currents up to 130 mA. (d) IMB of the ICL comb at ~19.288 GHz, as measured optically. (e) Far field image of the laser beam showing modal leakage to the device substrate, which is the likely cause of the irregularities in the optical spectrum.

### 2.2. RF conversion inside ICL observed in RIN and IMB

Sweeping the ICL's injection current revealed evolution of the optical spectrum as shown in Fig. 3(a), as well as the near-DC noise and RF IMB variabilities presented in Fig. 3(b) and (c), respectively. The latter two exhibit striking similarities in numerous regimes of the laser operation, with representative examples shown in Fig. 3(e)–(f). For instance, in regime III (66.5 mA), single-sideband modulation at multiples of ~2.1 MHz is observable in the vicinity of the IMB. We observe that the same modulation frequencies, rectified by the ICL, are also found in the low-frequency signal measured with a slow IR detector. Alternatively, the loss of modal coherence, which is visible as a flattened irregular IMB for 78.5 mA (regime V), is also seen at frequencies near zero. While some portion of the noise can be attributed to the LEF, the effects of frequency mixing are also noticeable. For instance, a characteristic bump in the RIN is clearly observable in the IMB spectrum as well. On the contrary, at 92.0 mA (regime VI), the beat note becomes relatively weak but remains sharp while the near-DC noise level increases. In this case, FM-to-AM noise conversion associated with the LEF dominates. It is highly probable that the newly-emerged group of modes at ~3110 cm$^{-1}$ contributes significant noise because it does not initially lock to the comb. After the dominant mode hop, which occurs after ~95 mA, this group has locked, and the RIN becomes quiet again. Interestingly, in regions where high mode coherence is expected, judging from the IMB (for instance, regime VII at 101.5 mA), the comb's near-DC noise becomes minimal. The other two current ranges that provide comparable comb stability are near 60 and 70 mA (regimes II and IV, respectively). Finally, in regime VIII, the IMB turns strong and broad, while the laser's near-DC RIN becomes significant and broadband. Here, the laser may have entered the liquid comb regime, although verification of that state would require an IMB-frequency-resolved SWIFTS measurement. It is essential to emphasize that a clean (almost noiseless) IMB always develops in low-RIN regimes. That makes the RIN spectrum a key diagnostic indicator of frequency comb stability.

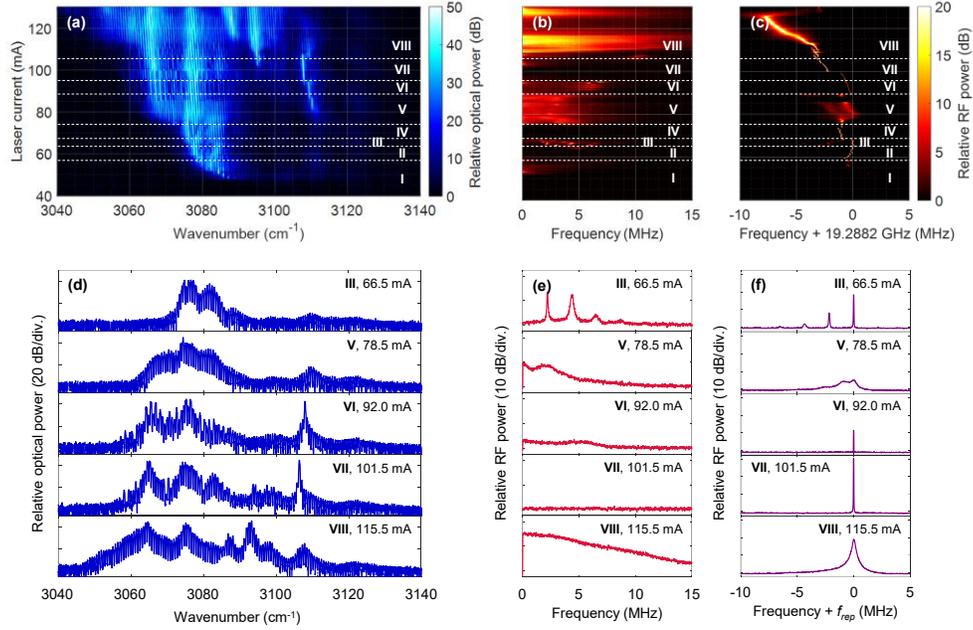

Fig. 3. (a) Optical spectra, (b) low-frequency RF noise spectra and (c) optical IMB spectra of the ICL output measured for injection currents ranging from 40 to 130 mA (Note: panels (a)-(c) share the same y-axis and panels (b)-(c) share the same color bar; capital roman numerals denote laser operation regimes). (d) Optical spectra, (e) low RF noise spectra, and (f) optical IMB spectra for selected values of the injection current (Note: the center frequency of the IMB $f_{rep}$ was determined independently for each spectrum).

The intrinsic frequency self-mixing observable in the measurements stems from the electrical nonlinearity of the ICL, which serves as a mixer with the local oscillator at $f_{rep}$, and also as a phase-to-amplitude noise converter caused by the linewidth enhancement factor (LEF). The latter phenomenon originates from the $\chi^{(3)}$ nonlinearity induced by the Bloch gain, which manifests experimentally as gain asymmetry [45]. Inside the laser, this nonlinearity converts modal phase (frequency) noise to amplitude noise. As a result, the baseband signal includes contributions from both electrical downconversion (self-mixing) and optical noise coupling (RIN). This is graphically explained in Fig. 1.

Capelli et al. proved experimentally that in semiconductor laser frequency combs, the frequency noise of an isolated comb line is the same as that of all lines, and is even lower than that of a single-mode laser [46]. This is possible because the FWM process effectively correlates the noise across all comb lines. However, when some lines are unlocked, their optical frequency noise becomes uncorrelated, and hence it cumulatively increases. This manifests in the electrical domain as increased RIN at low radio frequencies (baseband) caused by the FM-to-AM noise conversion. Our experiment illustrates this effect in regime VI, where the IMB is narrow and moderately strong (because many, but not all, of the modes are tightly locked), while the low-frequency noise is high and cannot be attributed solely to electrical downconversion. Spectral clustering or lasing on isolated mode groups separated in frequency promotes this effect. For instance, one sub-comb can form a sharp IMB, while another part of the spectrum is not locked at all, so the IMB signal lies below the noise floor. However, due to the lack of efficient FWM, intensity noise from all the comb lines (rather than equivalent to one) contributes to the low-frequency electrical spectrum. This makes it rich in RIN.

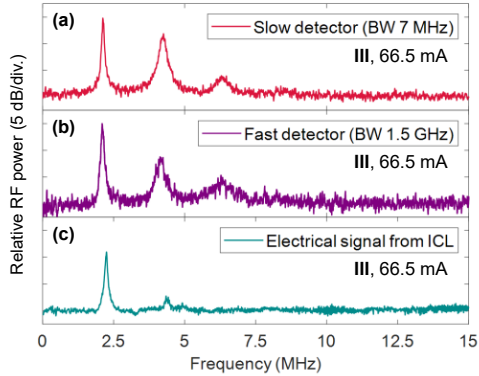

Fig. 4. Low RF noise spectra of the ICL operating at 66.5 mA, as obtained with a slow VIGO detector (a) and a fast ICIP (b), compared with the ICL's electrical spectrum extracted using a bias tee (c). Note: The spectra were not taken simultaneously, so small deviations in frequency and amplitude are expected due to the drift of the ICL's spontaneous modulation parameters.

To confirm that the observed frequency mixing is indeed intrinsic to the ICL and not attributable to the photodetectors (which are themselves electrically nonlinear components), an electrical spectrum was extracted directly from the laser using a bias tee. The results shown in Fig. 4 indicate that the near-DC noise of the ICL operating at 66.5 mA exhibits the same modulation sidebands on both photodetectors (Fig. 4(a) and (b)), as well as at the laser junction (Fig. 4(c)).

*2.3. Time-domain correlation between converted RF signals*

To more closely examine the temporal correlation of the original and converted modulation sidebands generated in the ICL at 66.5 mA, we simultaneously acquired the two signals in the time domain. The low RF signal from a VIGO Photonics detector was sampled by an oscilloscope (RIGOL Technologies, mod. DHO4204), while the fast IMB signal from an interband cascade photodetector was I/Q demodulated at the ICL's $f_{rep}$ by a real-time RF spectrum analyzer (Rohde&Schwarz, mod. FSW). Both recording devices had a sampling rate of 25 MS/s and were triggered externally by a 0.5 Hz rectangular waveform from the function generator with the same trigger level, and their clocks were all synchronized.

The electrical waveforms were recorded for 80 μs, with the middle 10% of traces shown in Figs. 5(a) and (b). As the conversion loss of the RF spectrum analyzer's built-in frequency mixer is similar to the strength of the strongest IMB sidebands, the magnitude of the converted in-phase and quadrature signals (tens of μV) is much smaller than the low RF signal measured by the slow detector (a few mV). Hence, for further analysis, the signals were detrended and normalized to values in the −0.5 to 0.5 range. Figure 5(b) presents their respective power spectral densities (PSDs) calculated from the entire 80-μs traces. Because the low RF signal is real, its PSD is symmetric. I/Q demodulation of the IMB, in contrast, yields a complex signal holding the single sideband modulation. A residual slow carrier peak is observed near zero in the PSD of the IMB, due to the imperfect match of the local oscillator to the ICL's $f_{rep}$. To confirm the temporal relation between the two signals, we employed the normalized cross-correlation function (CCF) plotted in Fig. 5(c). The initially low peak value of the CCF for raw samples (~0.07) can be increased significantly, to ~0.28, via filtering of the low RF and IMB signals using digital Gaussian bandpass filters at ±2.25 and ±4.5 MHz, and a full width at half maximum of 500 kHz.

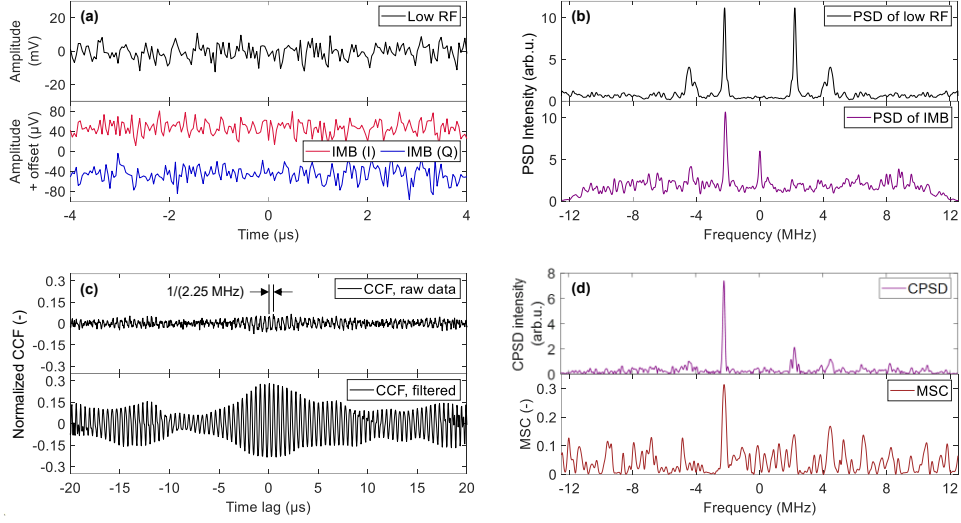

Fig. 5. (a) Low RF oscillations in the ICL output at 66.5 mA, as measured with a slow MCT detector (top), along with downconverted IMB oscillations (in-phase (I) and quadrature (Q) channels) measured with a fast ICIP (bottom). Note: A ±50 µV offset was added for each trace for clarity. (b) Power spectral density of the low RF (top) and downconverted IMB signals (bottom). Note: PSD of the complex real signal from the MCT is symmetric, while the PSD of IQ samples from the ICIP exhibits single-sideband modulation. The peak around zero is a residual carrier frequency. (c) Normalized cross correlation function (CCF) of the low RF and IMB IQ signals for raw samples (top), and after filtration of the signals (bottom; filter details are provided in the text). (d) Magnitude of the cross power spectral density (CPSD) of the low RF and IMB IQ signals (top) and magnitude-squared coherence (MSC) of the signals. Both show a strong common component at −2.25 MHz.

The correlation was also confirmed in the frequency domain by analyzing the cross-power spectral density (CPSD) – Fig. 5(d), top, and magnitude squared coherence (MSC) – Fig. 5(d), bottom. The CPSD is given by

$$S_{xy}(f) = \sum_{n=-\infty}^{\infty} R_{xy}[m] e^{-j2\pi fm}, \quad (1)$$

where the cross-covariance $R_{xy}$ is calculated as

$$R_{xy} = \sum_{n=-\infty}^{\infty} x[n] y^*[n-m]. \quad (2)$$

CPSD is a complex function with magnitude quantifying the correlation strength between two signals at a given frequency, and phase quantifying the relative delay. Here, they are obtained from the low-RF region and the IMB, respectively. For normalization purposes (to express the correlation between 0 and 1), the MSC normalizes the CPSD by a product of individual power spectral densities $S_{xx}$ and $S_{yy}$, being simply CPSDs for $y = x$ and $x = y$, respectively:

$$C_{xy}(f) = \frac{|S_{xy}(f)|^2}{S_{xx}(f) S_{yy}(f)}. \quad (3)$$

The peak value of 0.314 for the first MSC sideband proves the preserved coherence of the signals. With a better signal-to-noise ratio, especially for the demodulated IMB signal, a higher value of MSC would be expected. However, the MSC is not likely to reach unity because the modulation sidebands also include down-converted optical noise from frequency fluctuations due to the LEF.

## 3. Conclusion

In this work, we have demonstrated the internal conversion of radio frequency signals inside an ICL frequency comb. The device under test displayed efficient frequency downconversion from the vicinity of an IMB to near-DC noise. Correlations between these two bands were confirmed by analyzing spontaneous modulation signals in both the frequency and time domains. We identify the origin of the phenomenon as the electrical nonlinearity of the semiconductor laser, which forms a frequency self-mixer that internally converts the RF signals up and down. This, together with the linewidth enhancement factor responsible for phase-to-amplitude (frequency-to-amplitude/FM-AM) noise conversion, contributes to low-frequency electrical noise. Table 1 provides a convenient, practical summary that should aid in identifying the comb regimes given the observed low- and high-frequency electrical spectra. Note that although the first two regimes are characterized by low near-DC noise, they are easily distinguished by emission into either a single spectral mode or a comb-like spectrum. As in the latter group a quiet RIN spectrum was found exclusively in regimes where strong and sharp IMB was measured, it is possible to assess the comb stability based on near-DC spectrum only.

Table 1. Summary of observed laser regimes and their interpretations.

| Regime (Fig. 3) | Near-DC noise | IMB | Interpretation |
|---|---|---|---|
| I | Low | None | Quiet, single-mode operation |
| II, IV, VII | Low | Strong & sharp | Stable, locked frequency comb operation |
| III | Modulated | Modulated | Spontaneous modulation or coexistence of sub-combs |
| V | High | None or weak & broad | Unlocked state (the mode coherence is lost and broadband noise arises) |
| VI | Moderate | Weak & sharp | Partially-locked state (a sub-comb region is not locked and noisy, while the rest is locked) |
| VIII | High | Strong & broad | Liquid comb or unlocked state |

While in this work we focus on the microwave frequency downconversion from IMB to near-DC baseband, in principle the intracavity frequency conversion works in both directions. In particular, pronounced technical noise from the laser current driver, easily observable in RIN [47] at lower frequencies, can upconvert to the IMB via double sideband amplitude modulation (DSB-AM), which yields a symmetrical microwave spectrum. This, however, is in stark contrast to the typically observable sideband asymmetry around the IMB attributable to the lack of line equidistance in the optical spectrum. This phenomenon may produce a multi-peaked or wavy IMB spectrum with one dominant sideband. Empirically, we found frequency noise downconversion (from IMB to RIN – high-to-low frequencies) to dominate over upconversion when the laser is biased by a low-noise current driver.

Since the intrinsic frequency conversion mechanism is likely to be universal among semiconductor lasers, it creates a new and valuable diagnostic tool for assessing the stability of frequency comb operation. Our findings are expected to play a vital role in characterizing semiconductor laser devices in spectral regions where the availability of fast photodetectors is limited.

**Funding.** European Research Council (01117433).

**Acknowledgment.** G. Gomółka and Ł. A. Sterczewski acknowledge funding from the European Union (ERC Starting Grant, TeraERC, 101117433). Views and opinions expressed are, however, those of the authors only and do not necessarily reflect those of the European Union or the European Research Council Executive Agency. Neither the European Union nor the granting authority can be held responsible for them. NRL acknowledges support by ONR. This work is supported by use of the National Laboratory for Photonics and Quantum Technologies (NPLQT) infrastructure, which is financed by the European Funds under the Smart Growth Operational Programme.

**Disclosures.** The authors declare no conflicts of interest.

**Data availability.** The data that support the findings of this study are publicly available in Ref. [48].